# PERIODIC AND CHAOTIC EVENTS IN A DISCRETE MODEL OF LOGISTIC TYPE FOR THE COMPETITIVE INTERACTION OF TWO SPECIES

Ricardo LÓPEZ-RUIZ [*]
Danièle FOURNIER-PRUNARET [#]

[*] Department of Computer Science and BIFI,
Facultad de Ciencias-Edificio B,
Universidad de Zaragoza,
50009 - Zaragoza (Spain).

[#] Institut National des Sciences Appliquées,
Systèmes Dynamiques (SYD), L.E.S.I.A.,
Avenue de Rangueil, 31077 Toulouse Cedex (France).

## Abstract

Two symmetrically coupled logistic equations are proposed to mimic the competitive interaction between two species. The phenomena of coexistence, oscillations and chaos are present in this cubic discrete system. This work, together with two other similar ones recently published by the authors, completes a *triptych* dedicated to the two-species relationships present in Nature, namely the symbiosis, the predator-prey and the competition. These models can be used as basic ingredients to build up more complex interactions in the ecological networks.

# 1. COMPETITION ENFORCES BIODIVERSITY

Life is *folded* in DNA. When it *unfolds*, the individuals of the different species appear on the earth. Through them, DNA *readapts* to the new ambient conditions and passes to the next generation. Some species can become extinct, and the evolution pathway followed by its DNA finishes its *vital voyage* [1]. Surely, in the past, some ancestral evolutionary branching phenomenon on that DNA was the starting point of a new species. This species is now continuing its particular fight for existence. Hence, we can see the actual species as the many different *vehicles* that DNA is using to be present in the future.

An entangled set of relationships are established among all these species. It is evident that to reduce this complex ecological network to binary interactions between pairs of species, as for instance the *pairwise coevolution* [2], cannot solely explain the richness of the living world. Only the attempt of simplification can justify this way of approaching the problem. Thus, traditionally, three main types of interaction between individuals or species, namely the *mutualism*, the *predator-prey* situation or the *competition*, are considered as the basic bricks of *population dynamics* models [3]. These interaction schemes are not found in a pure way if not confounded among them in the majority of the ecosystems.

*Competition* is an undeniable force for evolution. When it is *intra-specific*, i.e., among individuals of the same species, it promotes a better adaptation to the environmental conditions. This continuous adaptation led Darwin [4] to conclude that *"the survivors are the fittest"*. This seems not to be a universal fact. Take, for instance, the case of human society, where the competition is sometimes perverted by many parameters of structural, economical, religious or sociological type in order to favour the most convenient individuals; those that are piloted by the *permanent* request of the local power [5]. In a more general context, if society is regarded as a whole, we can say that human laws are not established to reach a better stage for the collectivity by, for instance, sharing the wealth. At the contrary, they seem to be *imposed* in order to maintain the different economic classes present at each time [6]. Under this point of view, Pareto's law on the people's incomes could be interpreted as a simple irrational differentiation created by the richest economic classes; those that evidently manage the economic power. When competition is *inter-specific*, i.e., among different species, a diversification of the resource use is attained. If this is not the case, the strong dependence of the same limited resource for survival could derive in the extinction of some of the species. This possibility persuaded Gause [7] to enounce the *competitive exclusion principle*, which suggests that two species cannot coexist when they have identical need of a limited food. The evident consequence of this principle would be that more species than different resources cannot coexist in an ecosystem. Although the prototypical Lotka-Volterra differential equation model for two competing species [3] verifies this behavior in almost all circumstances, it also allows the coexistence when the competition between the species is weak. This last scenario confirms the reluctance to accept the competitive exclusion principle [8], which seems to be experimentally contradicted [9]. Therefore, it is reasonable to consider competition not necessarily as a destructive force but as an interaction that can generate interesting dynamical phenomena when the same resources are partitioned among several species. Thus, for instance, *oscillations and chaos* can be found in competitive systems [10, 11, 12]. Within this biological realm, competition is not a simple process, and the intrinsic complexity associated to this interaction allows and helps in the generation of *biodiversity*.

In this work, we propose a discrete logistic model for the competition between two species, which displays coexistence, oscillations, chaos and fractal behaviour of basins.

## 2. THE DISCRETE LOGISTIC MODEL FOR COMPETITION

### 2.1 The Logistic Model for One Isolated Species

If $x_n$ represents the population of an isolated species after $n$ generations, let us suppose that this variable is bounded in the range $0 < x_n < 1$. A simple model giving account of its evolution is the so called logistic map [11],

$$x_{n+1} = \mu x_n (1 - x_n), \qquad (1)$$

where $0 < \mu < 4$ in order to assure $0 < x_n < 1$. The term $\mu x_n$ controls the *activation or expanding phase*, with $\mu$ expressing the growth rate. The term $(1 - x_n)$ inhibits the overcrowding, and it controls *the inhibition or contracting phase*. The continuous version of this model was originally introduced by Verhulst in the nineteenth century as a counterpart to the Malthusian theories of human overpopulation.

When the growth rate is modified, the dynamical behaviour of the logistic equation (1) is the following:

(i) $0 < \mu < 1$: The growth rate is not big enough to stabilize the population. It will drop and the species will become extinct.

(ii) $1 < \mu < 3$: A drastic change is obtained when $\mu$ is greater than 1. The non vanishing equilibrium between the two competing forces, reproduction on one hand and resource limitation on the other, is now possible. The population reaches, independently of its initial conditions, a fixed value that is maintained in time.

(iii) $3 < \mu < 3.57$: A cascade of sudden changes provokes the oscillation of the population in cycles of period $2^n$, where $n$ increases from 1, when $\mu$ is close to 3, to infinity when $\mu$ is approaching the critical value 3.57. This is named the period-doubling cascade.

(iv) $3.57 < \mu < 3.82$: When the parameter moves, the system alternates between periodical behaviours with high periods on parameter interval windows and *chaotic regimes* for parameter values not located in intervals. The population can be not predictable although the system is deterministic. The chaotic regimes are observed for a given value of μ on sub-intervals of [0,1].

(v) $3.82 < \mu < 3.85$: The orbit of period 3 appears for $\mu = 3.82$ after a regime where unpredictable bursts, named *intermittences*, have become rarer until their disappearance in the three-periodic time signal. The existence of the period 3 orbit means, as the Sarkovskii theorem tells us, that all periods are possible for the population dynamics, although, in this case, they are not observable due to their instability. What is observed in this range is the period-doubling cascade $3*2^n$.

(vi) $3.85 < \mu < 4$: Chaotic behaviour with periodic windows is observed in this interval.

(vii) $\mu = 4$: The chaotic regime is obtained on the whole interval [0,1]. This specific regime produces dynamics, which looks like random. The dynamics has lost almost all its determinism and the population evolves as a random number generator.

Therefore, in addition to the rich birth process of its complex periodic orbit set, this system presents three remarkable dynamical behaviours: the period doubling route to chaos around the value $\mu \approx 3.57$ [13], the time signal complexification by intermittency in the neighbourhood of $\mu \approx 3.82$ [14] and the random-like dynamics for $\mu = 4$.

**2.2 The new Logistic Competition Model**

Let us suppose now two species $(x_n, y_n)$ that evolve according to a logistic-type dynamics. If they share the same ecosystem and they interact, we can approach their evolutions by the coupled system:

$$x_{n+1} = \mu_x(y_n)x_n(1-x_n),$$
$$y_{n+1} = \mu_y(x_n)y_n(1-y_n). \quad (2)$$

The interaction between both species causes the growth rate $\mu(z)$ to vary with time. Thus, $\mu(z)$ depends on the population size of the *others*. The simplest choice for this growth rate can be a linear increasing $\mu_1$ or decreasing $\mu_2$ function that expands the parameter interval in which the logistic map shows some activity, that is $\mu \in [1,4]$. Hence,

$$\mu_1(z) = \lambda_1(3z+1), \quad (3)$$
$$\mu_2(z) = \lambda_2(-3z+4). \quad (4)$$

Three different models are obtained for the three different relationships between species:

(i) The symbiosis originates a symmetrical coupling due to the mutual benefit, then $\mu_x = \mu_y = \mu_1$ [15].
(ii) The predator-prey interaction is based on the benefit/damage relationship established between the predator and the prey, respectively, then $\mu_x = \mu_1$ and $\mu_y = \mu_2$ [16].
(iii) The competition between species causes the contrary symmetrical coupling, then $\mu_x = \mu_y = \mu_2$.

These systems could be considered for future studies as the necessary bricks to build up more complex networks among interactive species. In particular, the model for the competition is

$$x_{n+1} = \lambda(-3y_n + 4)x_n(1-x_n),$$
$$y_{n+1} = \lambda(-3x_n + 4)y_n(1-y_n), \quad (5)$$

where $\lambda$ expresses the strength of the *mutual competitive interaction*. When some of the species is null the typical logistic behaviour is found for the other species.

The system (5) can be represented by the application $T_\lambda : [0,1] \times [0,1] \to \Re^2$, $T_\lambda(x_n, y_n) = (x_{n+1}, y_{n+1})$, where $\lambda$ is a positive constant which has sense in the range $0 < \lambda < 1.21$. In the following we shall write $T$ instead of $T_\lambda$ as the dependence on the parameter $\lambda$ is understood.

In Sections 3 and 4, we study more accurately the details of model (5). This represents a symmetric competition between two species that have access to the same amount of resources and that present a similar performance to compete for them. Other works in this direction exploit an asymmetrical relationship [17, 18] or a spatial distribution of the species [19], or the possibility of combining beneficial and detrimental effects in the mutual interaction between the species [20].

In order to summarize, we will explain first the dynamical behaviour of this coupled logistic system when $\lambda$ is modified. We obtain:

(i) $0 < \lambda < 0.25$: Although the competition is weak, the smallness of the growth rate provokes the extinction of species. We can also interpret that competition is an important force for survival in a pure competitive interaction and its absence or its weakness can cause extinction.

(ii) $0.25 < \lambda < 0.9811$: Both populations can survive and the system settles down in a stationary symmetrical equilibrium. This point is reached independently of the initial conditions. A greater $\lambda$ allows the coexistence of bigger populations.

(iii) $0.9811 < \lambda < 1.1743$: For $\lambda = 0.9811$ the destabilization of the fixed point gives rise to a stable period two cycle located on the diagonal. The populations oscillate now synchronously between the two values of that orbit.

(iv) $1.1743 < \lambda < 1.1875$: The period two cycle becomes unstable and an off-diagonal period four orbit appears. What could seem the starting point of a period-doubling cascade is just its final point. No new flip bifurcation is found for this particular orbit.

(v) $1.1875 < \lambda < 1.1924$: A Neimark-Hopf bifurcation happens and the period four orbit gives rise to a period four invariant closed curve. The pattern of the time behaviour of the populations is now quasi-periodic.

(vi) $1.1924 < \lambda < 1.201$: The dynamics becomes complex, then chaotic, and the populations visit successively in time an attractor which is formed by four chaotic rings.

(vii) $1.201 < \lambda < 1.206$: A crisis of the four chaotic bands with the diagonal changes its aspect, which is now formed by two big chaotic bands.

(viii) $\lambda > 1.206$: The iterations are going outwards the square $[0,1] \times [0,1]$ and evolve towards infinity. This critical value can be interpreted as some kind of catastrophe provoking the extinction of species.

Let us remark that competition is a constructive force in this model. A weak competition can lead the system to extinction. A stronger interaction can stabilize a fixed population or can also generate periodic or chaotic oscillations. Apparently there is no multi-stability in the coarse unfolding above presented but, as we show in section 4, a very localized zone in the parameter space, $1.1924 < \lambda < 1.1948$, is source of a fascinating dynamical richness; coexistence of high period orbits and their period doubling cascades succeed by slight modifications of the parameter $\lambda$. It reflects, if the biological simile is accepted, some kind of *explosion of life*. In a more general context of ecological networks, this fact puts in evidence that competition among species can enhance biodiversity, if this is interpreted as the different possibilities that Nature offers for the survival of species.

## 3. ATTRACTORS: NUMBER AND BIFURCATIONS

### 3.1 Table of Attractors

For the sake of clarity, firstly, the different parameter regions where the mapping $T$ has stable attractors in the interval $0 < \lambda < 1.206$ are given in the next table.

| INTERVAL | NUMBER OF ATTRACTORS | ATTRACTORS |
|---|---|---|
| $0 < \lambda < 0.25$ | 1 | $p_0$ |
| $0.25 < \lambda < 0.9811$ | 1 | $p_4$ |
| $0.9811 < \lambda < 1.1743$ | 1 | $(p_5, p_6)$ |
| $1.1743 < \lambda < 1.1875$ | 1 | $(p_{51}, p_{52}, p_{61}, p_{62})$ |
| $1.1875 < \lambda < 1.1924$ | 1 | period four invariant closed curve |
| $1.1924 < \lambda < 1.1948$ | multistability | high period orbits |
| $1.1948 < \lambda < 1.201$ | 1 | period four chaotic band |
| $1.201 < \lambda < 1.206$ | 1 | period two chaotic band |

The meaning of all these attractors is explained in the next sections.

### 3.2 Fixed Points, Period Two Cycle and Period Four Attractors

The fixed points, $(x_{n+1}, y_{n+1}) = T(x_n, y_n)$, of the map $T$ are located on the axes and on the diagonal. The restriction of $T$ over the axes is the logistic map $x_{n+1} = f(x_n)$, with $f(x) = 4\lambda x(1-x)$, and, similarly, the reduction of $T$ over the diagonal gives $f(x) = \lambda(4-3x)x(1-x)$. The solutions of $x_{n+1} = f(x_n)$ are $p_0, p_1, p_2$ on the axes and $p_3, p_4$ on the diagonal:

$$p_0 = (0,0),$$
$$p_1 = \left(\frac{4\lambda-1}{4\lambda}, 0\right), p_2 = \left(0, \frac{4\lambda-1}{4\lambda}\right),$$
$$p_3 = \frac{1}{6}\left\{7 + (1+12/\lambda)^{\frac{1}{2}}, 7 + (1+12/\lambda)^{\frac{1}{2}}\right\},$$
$$p_4 = \frac{1}{6}\left\{7 - (1+12/\lambda)^{\frac{1}{2}}, 7 - (1+12/\lambda)^{\frac{1}{2}}\right\}.$$

For $0 < \lambda < 0.25$, $p_0$ is an attractive node. For all the rest of parameter values, $p_0$ is a repelling node. The points $(p_1, p_2)$ exist for every parameter value and both are unstable for every value of $\lambda$. The point $p_3$ is located outside the square $[0,1] \times [0,1]$ and is also unstable for every value of $\lambda$. For $\lambda = 0.25$, $p_0$ and $p_4$ exchange stability in a transcritical bifurcation and $p_4$ becomes a stable node. When $\lambda = 0.9811$, $p_4$ loses stability by a flip bifurcation on the diagonal that gives rise to a period two orbit, $(p_5, p_6)$. This cycle becomes unstable by a transversal flip bifurcation that gives rise to an off-diagonal period four orbit, $(p_{51}, p_{52}, p_{61}, p_{62})$, for $\lambda = 1.1743$. For this critical value, $p_5 = (0.33, 0.33)$ and $p_6 = (0.78, 0.78)$. When $\lambda = 1.1875$ the off-diagonal period four orbit, with $p_{51} = (0.34, 0.16)$, $p_{52} = (0.16, 0.34)$, $p_{61} = (0.47, 0.94)$ and $p_{62} = (0.94, 0.47)$, destabilizes by a Neimark-Hopf bifurcation towards a period four invariant curve. A rich sequence of bifurcations, which are described in the next section, can be detected in a narrow range of the parameter interval when this invariant closed curve transits towards a period four chaotic band.

## 4. CHAOTIC EVENTS AND FRACTAL BASINS

Here we try to enlighten what kind of bifurcations happens in the parameter interval $1.1924 < \lambda < 1.1948$. Due to the richness of bifurcations occurring in this small parameter region we can only give some particular examples of this great dynamical complexity. Also how the different initial populations evolve towards their asymptotic stable state is studied. In our mathematical representation, this is exactly the problem of considering the *basins* of the different attractors of model (5). For the sake of coherence, we will consider the square $[0,1] \times [0,1]$ as the source of initial conditions having sense in our biological model, i.e., in the map $T$. Let us say at this point that basins constitute an interesting object of study by themselves. If a colour is given to the basin of each attractor, we obtain a coloured figure, which is a phase-plane visual representation of the asymptotic behaviour of the points of interest. The strong dependence on the parameters of this coloured figure generates a rich variety of complex patterns on the plane and gives rise to different types of basin fractalization. See, for instance, the work done by Gardini *et al.* [21] and by López-Ruiz & Fournier-Prunaret [22, 23] with coupled logistic mappings.

### 4.1 Definitions and General Properties of Basins

The set $D$ of initial conditions that converges towards an attractor at finite distance when the number of iterations of $T$ tends towards infinity is the basin of the attracting set at finite distance. When only one attractor exists at finite distance, $D$ is the basin of this attractor. When several attractors at finite distance exist, $D$ is the union of the basins of each attractor. The set $D$ is invariant under backward iteration $T^{-1}$ but not necessarily invariant by $T$: $T^{-1}(D) = D$ and $T(D) \subseteq D$. A basin may be connected or non-connected. A connected basin may be simply connected or multiply connected, which means connected with holes. A non-connected basin consists of a finite or infinite number of connected components, which may be simply or multiply connected. The closure of $D$ also includes the points of the boundary $\partial D$, whose sequences of images are also bounded and lay on the boundary itself. If we consider the points at infinite distance as an attractor, its basin $D_\infty$ is the complement of the closure of $D$. When $D$ is multiply connected, $D_\infty$ is non-connected, the holes (called lakes) of $D$ being the non-connected parts (islands) of $D_\infty$. Inversely, non-connected parts (islands) of $D$ are holes of $D_\infty$ [24, 25].

In Sec. 3, we explained that the map (5) possesses, in general, one attractor at a finite distance. The points at infinity constitute the second attractor of $T$. In the region of our interest $1.1924 < \lambda < 1.1948$, multistability is possible. Thus, if we give a different colour for the basin of each attractor, a coloured pattern in the square $[0,1] \times [0,1]$ is obtained. In the present case, the phenomena of finite basin fractalization and its disappearance have their origin in the competition between the attractors at finite distance (whose basin is $D$). When a bifurcation of $D$ takes place, some important changes appear in the coloured figure, and, although the dynamical causes cannot be clear, the coloured pattern becomes an important visual tool to analyze the behaviour of basins (see [25] for further details).

The map $T$ is said to be noninvertible if there exist points in state space that do not have a unique rank-one preimage under the map. The state space is divided into regions, named $Z_i$, in which points have $i$ rank-one preimages under $T$. These regions are separated by the so called critical $LC$ curves, which are the images of the $LC_{-1}$ curves: $LC = T(LC_{-1})$. If the map $T$ is continuous and differentiable, the $LC_{-1}$ curve is the locus of points where the determinant of the Jacobean matrix of $T$ vanishes [25]. For the map (5), $LC_{-1}$ is formed by the points $(x, y)$ that satisfy the equation:

$$27x^2 y^2 - 57x^2 y - 57xy^2 + 24x^2 + 24y^2 + 112xy - 44x - 44y + 16 = 0. \quad (6)$$

$LC_{-1}$ is independent of $\lambda$ parameter and is quadratic in $x$ and $y$. It is a curve of four branches, with two horizontal and two vertical asymptotes. $LC$ depends on $\lambda$ and separates the plane into three regions that are locus of points having 1, 3 or 5 distinct preimages of rank-1. They are denoted by $Z_i$, $i = 1,3,5$, respectively. The region $Z_3$ is not connected and it is formed by five disconnected zones in the plane. It is straightforward to verify the number of preimages of each one of the $Z_i$ zones by directly calculating this number for different points located in each one of the five intervals on the diagonal in which is divided by the $LC$ curves. A detailed calculation of a similar problem has been performed in [15, 16]. Thus, the point $(0,0)$ has five preimages, which are $\{(0,0),(0,1),(1,0),(1,1),(1.33,1.33)\}$, independently of $\lambda$. For $\lambda = 1$, the point $(-1,-1)$ has three preimages which are $\{(1.13,-0.88),(-0.88,1.13),(-0.18,-0.18)\}$, the point $(0.5,0.5)$ presents also three preimages which are $\{(0.17,0.17),(0.62,0.62),(1.53,1.53)\}$, the point $(1,1)$ has an only preimage which is $\{(1.64,1.64)\}$ and the point $(3,3)$ presents three preimages which are $\{(1.57,3.17),(2.07,2.07),(3.17,1.57)\}$. When $\lambda$ is modified, the pattern of the $Z_i$ zones is maintained. According to the nomenclature established in Mira *et al.* [25], the map (5) is of type $Z_3 - Z_5 \succ Z_3 - Z_1 \prec Z_3$. The evolution of critical manifolds regarding the modification of the parameter can explain the bifurcations of the basins, for instance the appearance of holes inside a basin or the fractalization of basin boundaries [22, 25].

### 4.2 High Period Orbits, Chaotic Bands and their Basins

The region of the parameter space $1.1924 < \lambda < 1.1948$ is now investigated. Some examples of the dynamical richness of this region are given in the next table. High period orbits (PO), invariant closed curves (ICC), weakly chaotic rings (WCR) and chaotic bands (CB) are created by slight modifications of $\lambda$. The unfolding of period doubling cascades of some high period orbits can be accurately discerned. We observe a signature that characterizes this kind of successive bifurcations; the final state of the cascade is a chaotic band formed by a number of so many small disconnected pieces as the initial period of the orbit, which gave rise to the period doubling cascade. For instance, we have observed the complete period doubling cascade of a period $4\times 19$ orbit in the interval $1.19385 < \lambda < 1.19389$ and similarly the complete cascade of a pair of period $4\times 13$ orbits in the interval $1.1943 < \lambda < 1.19445$ (see Figures 6-7-8).

Multistability among several attractors is also possible. As it can be seen in the figures, the basins in the square $[0,1]\times[0,1]$ take complex and fractal forms. Some of them are riddled basins, that is, successive zooms of a basin zone cannot delimitate the boundaries between the basins of the different attractors. A resume of our observations is given in the next table.

| $\lambda$ | NUMBER OF ATTRACTORS | ATTRACTORS | FIGURES |
|---|---|---|---|
| 1.192693 | 2 | two $4\times 11$ PO | 1 (a-b) |
| 1.19288 | 4 | two $4\times 3$ PO, one $4\times 11$ PO, one $4\times 33$ PO | 2 (a-b) |
| 1.1932 | 3 | one $4\times 11$ PO, two $4\times 3$ ICC | 3 (a-b) |
| 1.19342 | 3 | one $4\times 17$ PO, two $4\times 3$ WCR | 4 (a-b) |
| 1.19345 | 2 | two 4x17 PO | 5 (a-b) |
| 1.1935 | 1 | one $4\times 1$ CB | 6 (a-b) |
| 1.1941 | 2 | two $4\times 19$ PO | 7 (a-b) |
| 1.19435 | 2 | two $4\times 13$ PO | 8 (a-b-c-d) |
| 1.1946 | 2 | two $4\times 13$ WCR | 9 (a-b-c-d) |
| 1.1948 | 1 | one $4\times 1$ CB | 10 |
| 1.201 | 1 | one $2\times 1$ CB | 11 |
| 1.205 | 1 | one $2\times 1$ CB | 12 (a-b) |

The origin of all these attractors is the off-diagonal period four orbit of section 3.2. All the zooms made in these figures are just done on the neighbourhood of one of the four points which compose the period four orbit. This is the reason of the factor 4 that appears in all the orbits of the table above.

### 4.3 Fractal Structure of the Basin Boundary

The fractal structure that is observed in the boundary of the basin (Fig. 13) can be explained by regarding the preimages of the origin, $p_0 = (0,0)$. This unstable fixed point belongs to the closure of the stable manifold of $p_4$ for $0.25 < \lambda < 0.9811$ and to the closure of the stable manifold of the period two cycle $(p_5, p_6)$ for $\lambda > 0.9811$. (These stable manifolds are mostly located on the diagonal of the square $[0,1]\times[0,1]$. Let us recall that for $\lambda > 1.1743$ the diagonal loses its transversal stability by a transversal flip bifurcation of the period two cycle $(p_5, p_6)$). Therefore, the origin and all its preimages belong to the boundary of the basin. Hence the points where the basin contacts the axes are just preimages of the origin. These points, which have coordinates $(z_n, 0)$ and $(0, z_n)$, are obtained by pairs with the double recurrence

$$z_{n+1} = \frac{1}{2}(1 \pm \sqrt{1 - z_n / \lambda}), \quad (6)$$

when it is fed with the initial condition $z_0 = 1$. This equation is obtained by inverting the logistic map commanding the dynamics on the axes. The initial condition $z_0 = 1$ is necessary because the extreme points $(1,0)$ and $(0,1)$ are the first rank preimages of the origin.

The boundary basin points with coordinates $(\overline{z_n},1)$ and $(1,\overline{z_n})$ are calculated from the $z_n$ points obtained from the recurrence (6) by the expression

$$\overline{z_n} = \frac{1}{2}(1 \pm \sqrt{1 - 4z_n/\lambda}). \qquad (7)$$

This expression is obtained by observing that the image of a point with coordinates $(\overline{z_n},1)$ or $(1,\overline{z_n})$ is a point with the form $(z_n,0)$ or $(0,z_n)$, respectively, and that they are related by a logistic equation. Evidently this expression has only sense when $z_n < \lambda/4$. Therefore, the fractal contact set between the basin and the axes can be exactly obtained by iterating expressions (6-7) with the initial condition $z_0 = 1$ (Fig. 13).

## 5. CONCLUSIONS

Three main types of interaction between two species, namely the predator-prey situation, competition or mutualism, are usually taken as basic bricks for understanding how the populations evolve in an ecosystem. A first theoretical approach to model these interactions is the well-known Lotka-Volterra equation [3] and all its different versions. These quadratic differential equations can display extinction or coexistence between species for the three types of interaction. Periodic behaviour is also possible for the predator-prey case, but multistability or chaotic oscillations are not present in these equations. Nevertheless, this type of behaviour has been obtained, for instance, when a competitive interaction among species takes place [9, 11]. Some ingredient would be missing in the Lotka-Volterra models in order to have a chaotic dynamics.

We have undertaken a different approach in order to have chaotic oscillations in our models. We suppose that the growth rate of the species is not independent of the number of individuals of the other species. It is linearly modified with that quantity. The result is a group of three cubic two-dimensional coupled logistic equations which are proposed to mimic the three different types of two species relationships. Two of them have been studied in the cadre of mutualism and predator-prey [15, 16]. In this work we have carried out the study of the third system for the case of two competing species. Although the interaction has been supposed to be symmetrical, the system displays chaotic dynamics. In fact, there exists a region in the parameter space where the system shows a rich structure of bifurcations. High period orbits and their period doubling cascades have been identified, their basins have been calculated and their fractal structures have been enlightened.

In the same way that ecological interactions has fitness consequences for species, the possibility to have new simple views, although unrealistic, on the basic interactions between species can enlarge the spectrum of combinations for building up more complex images of the ecosystems. Following this path and using some of these new ingredients, some models with different number of elements have recently appeared in the literature [26, 27, 28]. We are sure that these works can help for increasing our insight about ecological networks.

**Acknowledgements**

We would like to thank Dr. Taha for his helpful comments. R. L.-R. wishes also to thank the *Systèmes Dynamiques* Group at LESIA-INSA (Toulouse) for its kind hospitality and M. Hernández-Pardo for useful comments, and D. F.-P. thanks the Faculty of Sciences of the Universidad de Zaragoza. This work was supported by the Spanish MCYT projects HF2002-0076 and FIS2004-05073-C04-01, and by the French research project EGIDE-PICASSO 05125VC.

# **Figure Captions**

**Fig. 1:** Basin of two $4\times 11$ PO.
**Fig. 2:** Basin of two $4\times 3$ PO, one $4\times 11$ PO and one $4\times 33$ PO.
**Fig. 3:** Basin of one $4\times 11$ PO and two $4\times 3$ ICC.
**Fig. 4:** Basin of one $4\times 17$ PO and two $4\times 3$ WCR
**Fig. 5:** Basin of two 4x17 PO. These are riddled basins.
**Fig. 6:** Basin of one $4\times 1$ CB.
**Fig. 7:** Basin of two $4\times 19$ PO.
**Fig. 8:** Basin of two $4\times 13$ PO. Successive enlargements of these riddled basins are given.
**Fig. 9:** Basin of two $4\times 13$ WCR.
**Fig. 10:** Basin of one $4\times 1$ CB.
**Fig. 11:** Basin of one $2\times 1$ CB.
**Fig. 12:** Basin of one $2\times 1$ CB before than it disappears by a contact bifurcation with its basin boundary.
**Fig. 13:** Details of the basin boundary on the axes; successive enlargements showing the points $(z_n,0)$, which are preimages of the origin, are given.

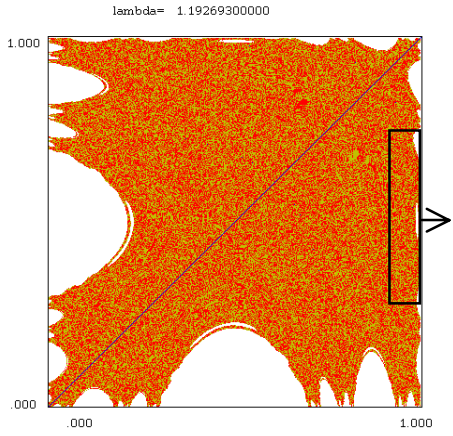

Fig. 1a

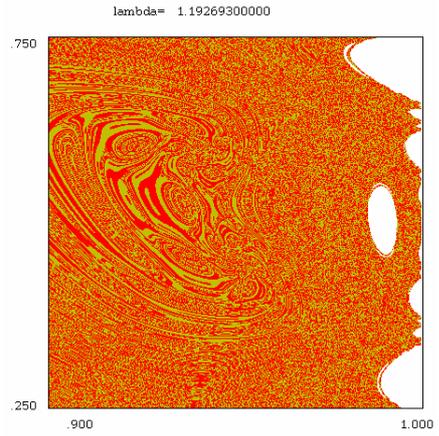

Fig. 1b

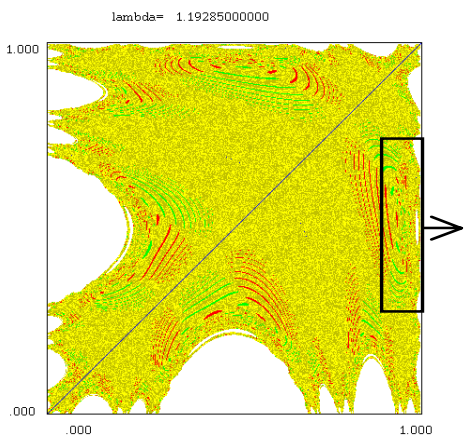

Fig. 2a.

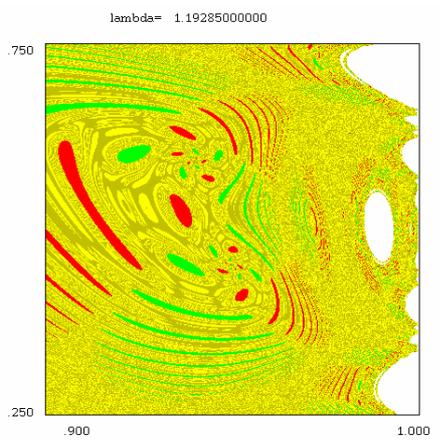

Fig. 2b

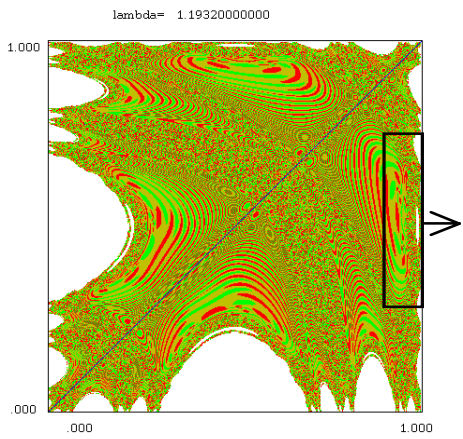

Fig. 3a

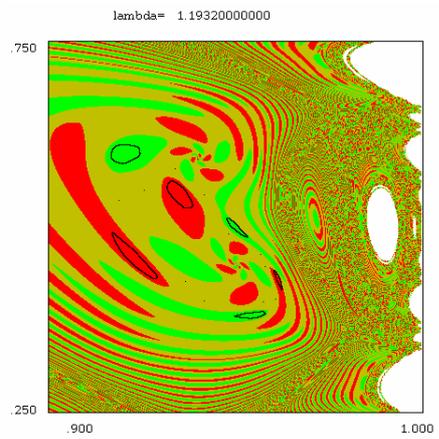

Fig. 3b

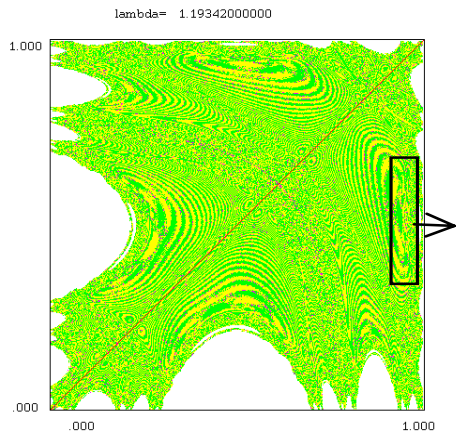

Fig. 4a

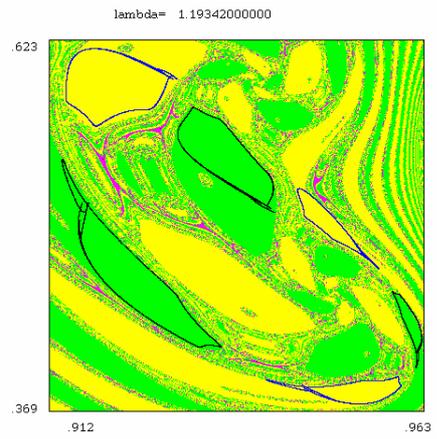

Fig. 4b

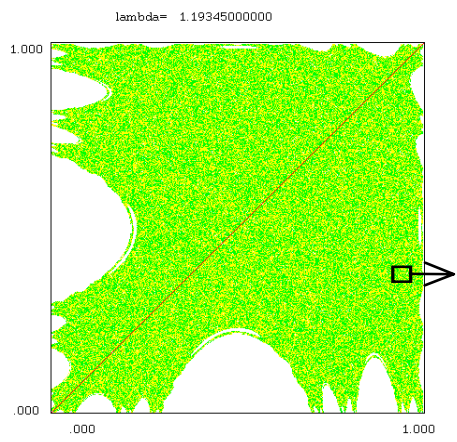

Fig. 5a

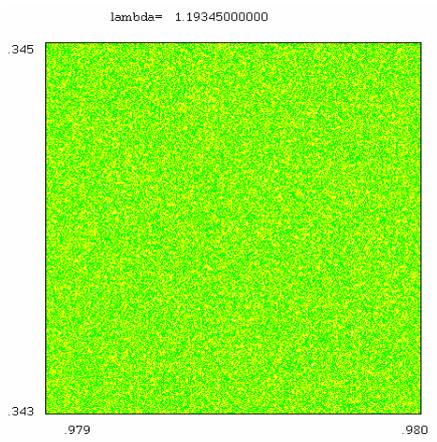

Fig. 5b

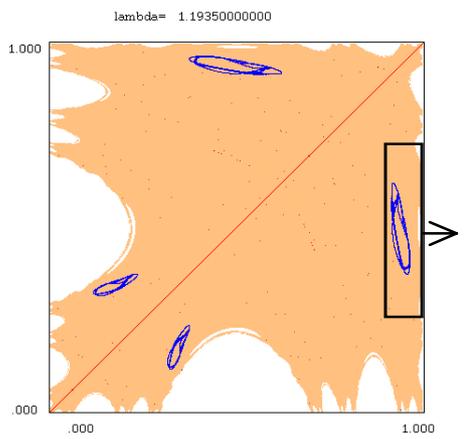

Fig. 6a

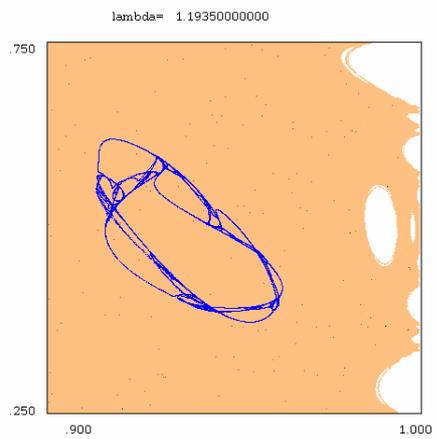

Fig. 6b

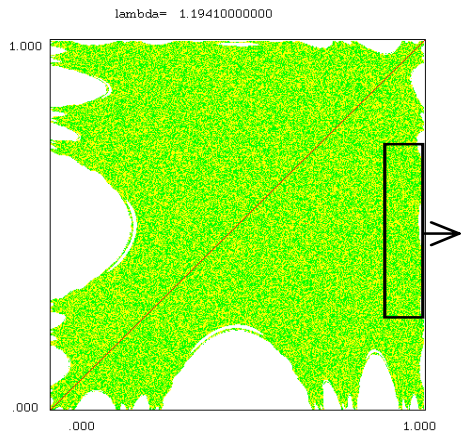
Fig. 7a

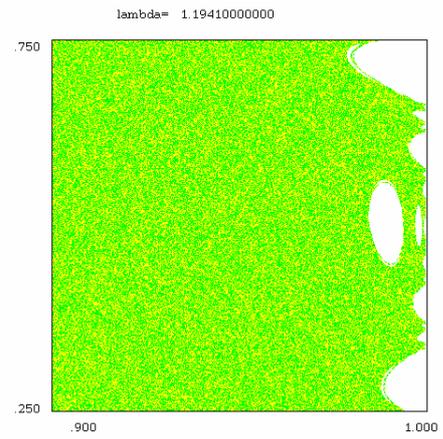
Fig. 7b

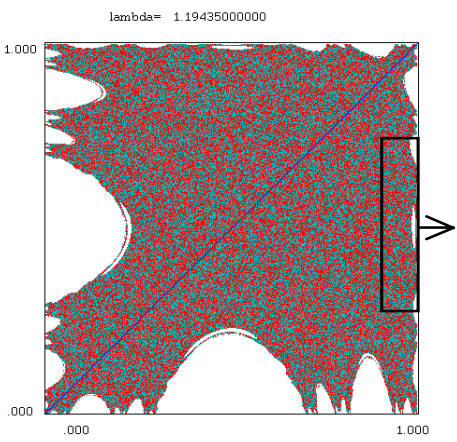
Fig. 8a

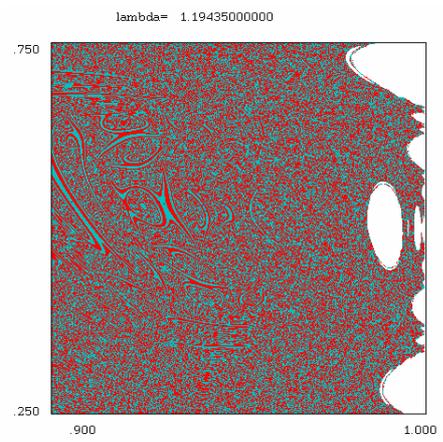
Fig. 8b

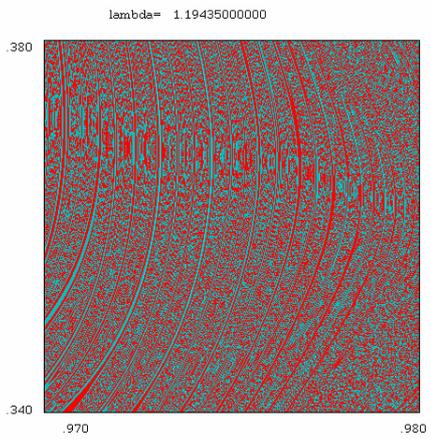
Fig. 8c

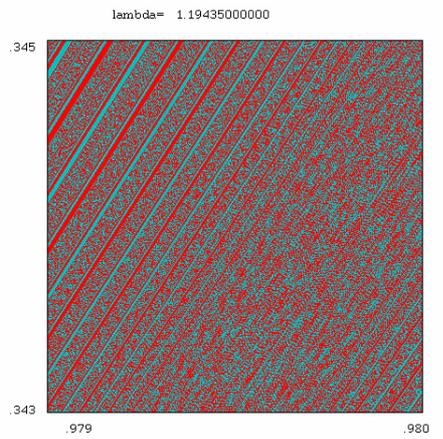
Fig. 8d

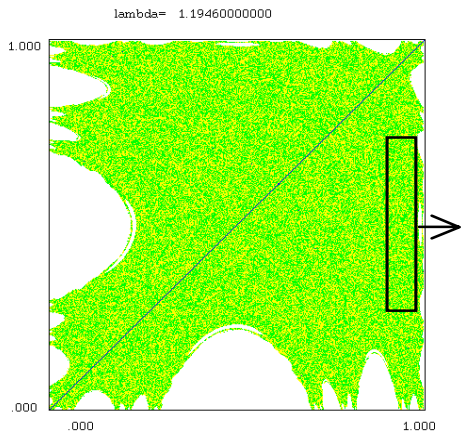

Fig. 9a

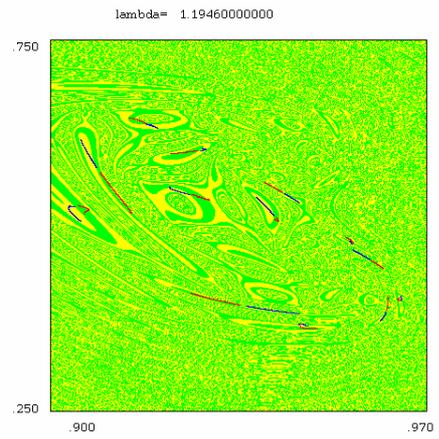

Fig. 9b

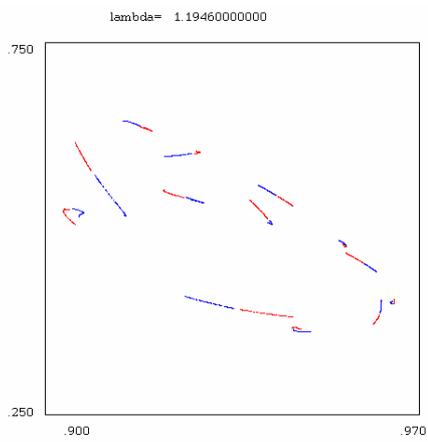

Fig. 9c

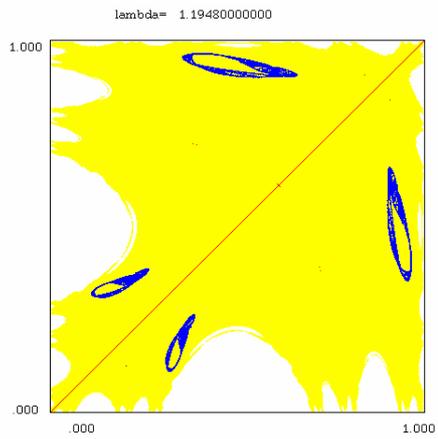

Fig. 10

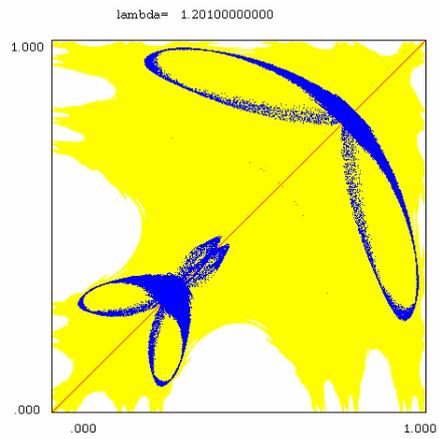

Fig. 11

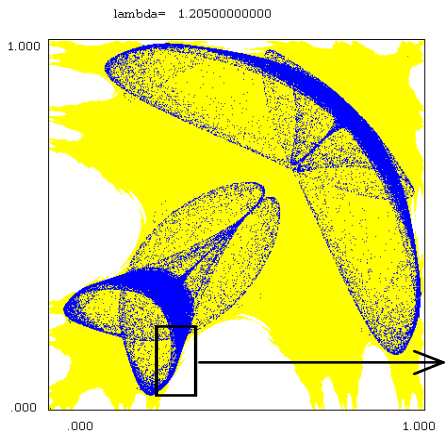

Fig. 12a

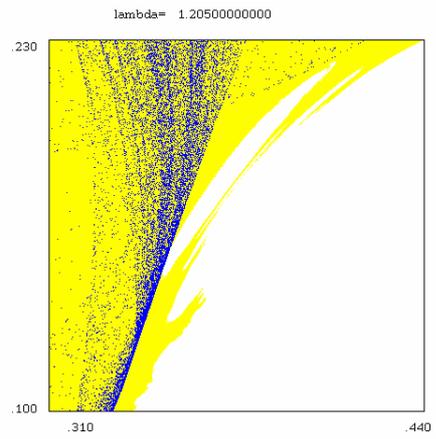

Fig. 12b

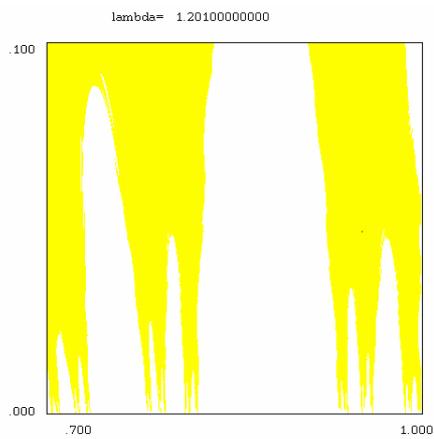

Fig. 13a

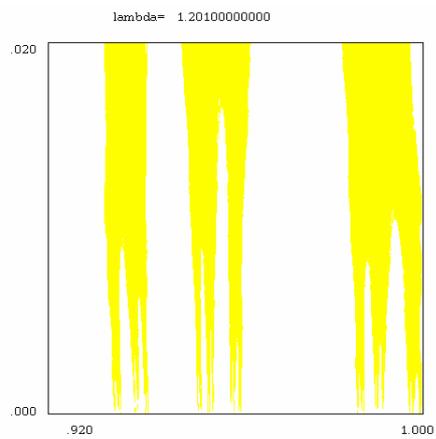

Fig. 13b

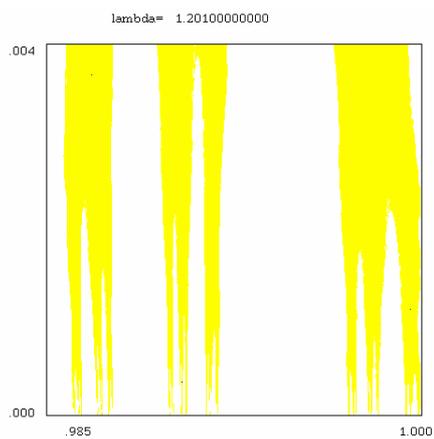

Fig. 13c